# Surface energy coefficient determination in global mass formula from fission barrier energy

Serkan Akkoyun[1,*] and Tuncay Bayram[2]

[1]*Cumhuriyet University, Faculty of Science, Department of Physics, Sivas, Turkey*

[2]*Sinop University, Department of Nuclear Energy Engineering, Sinop, Turkey*

**Abstract**

Semi-empirical mass formula of the atomic nucleus describe binding energies of the nuclei. In the simple form of this formula, there are five terms related to the properties of the nuclear structure. The coefficients in each terms can be determined by various approach such as fitting on experimental binding energy values. In this study, the surface energy coefficient in the formula which is a correction on total binding energy has been obtained by a method that is not previously described in the literature. The experimental fission barrier energies of nuclei have been used for this task. According to the results, surface energy coefficient in one of the most conventional formula has been improved by a factor 3.4.

## 1. Introduction

The nuclear mass formula is very important for describing nuclear properties and exploring the exotic structure of the nuclei such as halo structure, super-heavy nuclei structures and decays [1]. Liquid drop model clarifies many nuclear phenomena which are unachievable by the shell model of the nucleus. The semi-empirical mass formula based on this model of the nucleus was first proposed in 1935 by Bethe and von Weizsacker [2, 3]. According to the formula, the nuclear binding energy is expressed in terms of A and Z numbers of the nuclei. The conventional formula has simply five terms named as volume, surface, Coulomb, asymmetry and pairing energy terms. The surface term is a correction in total binding energy due to deficit of binding energy for nucleons in the surface area. The magnitude of the nuclear surface energy is intimately related to the diffuseness of the nuclear surface and should provide a measure of the thickness of the nuclear surface. Since the surface energy is related to the lack of binding of the particles in the surface, it is clear that any attempt at a quantitative account of the nuclear surface energy will come up against difficulties due to our insufficient understanding of the nature of the effects responsible for nuclear cohesion [4]. Besides, the well-known force inside

Corresponding author e-mail: serkan.akkoyun@gmail.com

the nucleus is related to Coulomb energy term. This is a repulsive term among the protons. The coefficient in the Coulomb term can easily be calculated by the formula $a_c = 3e^2/5r_0$. Recently, semi-empirical mass formula has been extended by adding extra terms or has been modified slightly or completely [5-13]. These attempts have been made in order to obtain binding energies of the nuclei more accurately. In the study of Kim and Cha [14], the coefficients and even the power of the A number have been determined in order to reach experimental values as close as possible. Also in that work, the nuclei are divided into different groups according to their half-lives and it is obtained different coefficients for each group. The coefficients in each term can be determined by fitting the formula to the experimental binding energies on the atomic nuclei.

After the discovery of the fission, this phenomenon was started studying by considering nuclear drop model. If the Coulomb energy does not exceed a critical value, a charged drop is stable against fission. The surface energy in the drop model wants to keep the nucleus spherical, whereas Coulomb energy wants to deform it. Whether there will be a fission or not depends on the balance of these two effects. One can determine fissility parameter (*x*) that is characterized by the ratio of surface and Coulomb energies. If *x* exceeds the value of 1, fission occurs immediately [15].

Throughout the years, the constant in the semi-empirical mass formula has been determined many times by using various procedures or on different data sets. Every determined coefficient is different from each other. In this study, we have applied a different approach to obtain a constant in the basic five term formula. We have used experimental fission barrier energies to determine the surface energy coefficient in semi-empirical mass formula. We have taken the measured fission barriers from Myers study [16]. We have considered *x* to perform this task. Our aim was to obtain surface energy coefficient from experimental fission barrier energies and hence to reduce the mean square error value between theoretically determined binding energies of the nuclei and experimental ones.

## 2. Material and Methods

The most conventional simple semi-empirical mass formula considered in this work has been given in Eq. 2.1. This formula has simply five terms named as volume, surface, Coulomb, asymmetry and pairing. The coefficients in each term are calculated mostly by fitting to

Corresponding author e-mail: serkan.akkoyun@gmail.com

experimentally measured masses of nuclei. They usually vary depending on the fitting methodology.

$$B \ (MeV) = a_v A - a_s A^{2/3} - a_C \frac{Z(Z-1)}{A^{1/3}} - a_a \frac{(A-2Z)^2}{A} + a_p \frac{k}{A^{3/4}} \qquad (2.1)$$

In the formula, $k$ takes the value +1, 0 or -1 for even-even, even-odd or odd-odd nuclei, respectively.

In fission process in which nuclear shape deviates from spherical shape, the surface energy of the nuclei increases and the Coulomb energy decreases because charge density is reduced. The other terms contributing to the total binding energy of the nuclei are not appreciable changed when the nuclei split into two fragments. The total potential energy is determined by the sum of these two terms given in Eqs. 2.3-4.

$$E_C^0 = a_c \frac{Z(Z-1)}{A^{1/3}} \qquad (2.3)$$

$$E_s^0 = a_s A^{2/3} \qquad (2.4)$$

The ratio of these terms given in Eq. 2.5 has been known as fissility parameter $x$. Stable, unstable, and metastable states are defined using the fissility parameter, the released energy, and the fission barrier [16].

$$x = \frac{E_C^0}{2E_s^0} = \frac{a_C}{2a_s} \frac{Z(Z-1)}{A} \qquad (2.5)$$

Here, $E_C^0$ and $E_s^0$ are Coulomb and surface energies of the spherical nucleus. If the changes in the Coulomb and surface energies are equal to each other according to their spherical states, the nucleus becomes unstable against fission. This parameter is reached to 1 for *Z(Z-1)/A≈50*. Hence, according to the drop model of the nucleus, nuclei with *Z(Z-1)/A>50* are unstable against fission [17].

The liquid drop model of the nucleus permits calculation of the change in potential energy of the nucleus when it deviates from spherical shape [18]. In this case, the potential energy of the nucleus increases. The contributions to this change comes from surface and Coulomb energy terms. The Coulomb energy repulsion wants to deform spherical shape while the surface energy wants to keep nucleus spherical. The total change in potential energy is the total deformation energy and it is considered as in Eq. 2.6.

$$\Delta E = (E_s + E_C) - (E_s^0 + E_C^0) = E_s^0 [\frac{2}{5}(1-x)a_2^2 - \frac{4}{105}(1+2x)a_2^3 + \cdots \qquad (2.6)$$

Corresponding author e-mail: serkan.akkoyun@gmail.com

where $E_s$ and $E_C$ are surface and Coulomb energy of the deformed nucleus, $a_2 = (5/4\pi)^{1/2}\beta_2$. We can calculate the maximum of Eq.2.6 as

$$\frac{d\Delta E}{da_2} = 0 = E_s^0 [\frac{4}{5}(1-x)a_2 - \frac{4}{35}(1+2x)a_2^2] \qquad (2.7)$$

The first root ($a_2 = 0$) corresponds to minimum of the spherical nucleus and the second ($a_2 = 7(1-x)/(1+2x)$) is fission barrier maximum. If we substitute the second root to Eq.2.6 we can obtain fission barrier maximum in MeV. The fission barrier maximum is determined as difference between the saddle-point and ground state masses. This can be calculated theoretically by using Eq 2.8 as

$$E_b = \frac{98.(1-x)^3}{15.(1+2x)^2} \cdot E_s^0 \qquad (2.8)$$

where $E_s^0$ and $E_b$ are surface energies of the spherical nucleus and barrier energy. If experimental barrier energies of the fissionability nuclei are used in this formula and by considering the $E_C^0$ is well-known, it can be confidently calculated the surface energies $E_s^0$ of the nuclei. After obtaining this energy values for different nuclei, it is easy to have surface energy coefficient $a_s$.

### 3. Results and Discussion

We have used the barrier maximum formula (Eq. 3.1) in order to obtain surface term coefficient in the semi-empirical mass formula.

$$E_b = \frac{98.(1-\frac{E_C^0}{2E_s^0})^3}{15.(1+\frac{E_C^0}{E_s^0})^2} \cdot E_s^0 \qquad (3.1)$$

where $E_b$, $E_C^0$ and $E_s^0$ are maximum energy of fission barrier, Coulomb energy and surface energy for spherical nuclei, respectively. By solving this qubic equation, we have obtained surface energy ($E_s^0$) of the nuclei. We have considered Eq. 2.3 for Coulomb energy and taken the coefficient, $a_C = 0.72$, as given in the coefficient from Krane [19]. We have thought that if one can take any experimental values to derive something, this procedure can be one of the best way for this aim. Therefore, we have used experimental fission barrier height in MeV [20]. This data file includes total 36 isotopes of the nuclei from *Lu* (Z=71) to *Cf* (Z=98). After determination of $E_s^0$ by Eq. 3.1, we have used Eq. 2.4 to get surface term coefficient. As can be seen in Table 1 that the surface term coefficients have been calculated for different isotopes

Corresponding author e-mail: serkan.akkoyun@gmail.com

which have experimental barrier data. From all 36 isotopes, we have calculated the average value of the coefficient. According to the results, the coefficient has been redefined as 16.481.

**Table 1.** Measured fission barrier [20], Coulomb and surface energies and surface term coefficient for 36 isotopes.

| Z | N | A | $E_b$ | $E_C^0$ | $E_s^0$ | $a_s$ |
|---|---|---|---|---|---|---|
| 71 | 102 | 173 | 28.00 | 642.2048 | 498.428 | 16.054 |
| 73 | 106 | 179 | 26.10 | 671.4858 | 513.417 | 16.165 |
| 75 | 110 | 185 | 24.00 | 701.2963 | 527.681 | 16.253 |
| 76 | 110 | 186 | 23.40 | 718.9572 | 537.645 | 16.500 |
| 76 | 111 | 187 | 22.70 | 717.6733 | 534.824 | 16.355 |
| 76 | 112 | 188 | 24.20 | 716.3986 | 538.180 | 16.399 |
| 77 | 112 | 189 | 22.60 | 734.2031 | 545.352 | 16.559 |
| 77 | 114 | 191 | 23.70 | 731.6314 | 546.807 | 16.487 |
| 80 | 118 | 198 | 20.40 | 780.7186 | 568.856 | 16.745 |
| 81 | 120 | 201 | 22.30 | 796.4811 | 585.002 | 17.049 |
| 83 | 124 | 207 | 21.90 | 828.3890 | 604.401 | 17.272 |
| 83 | 126 | 209 | 23.30 | 825.7381 | 607.044 | 17.237 |
| 84 | 126 | 210 | 20.95 | 844.5333 | 611.723 | 17.314 |
| 84 | 128 | 212 | 19.50 | 841.8691 | 605.130 | 17.020 |
| 85 | 128 | 213 | 17.00 | 860.8038 | 608.137 | 17.051 |
| 88 | 140 | 228 | 8.10 | 902.3108 | 591.541 | 15.850 |
| 90 | 138 | 228 | 6.50 | 944.0320 | 605.706 | 16.230 |
| 90 | 140 | 230 | 7.00 | 941.2877 | 607.629 | 16.187 |
| 90 | 142 | 232 | 6.30 | 938.5751 | 601.014 | 15.918 |
| 90 | 144 | 234 | 6.65 | 935.8934 | 601.968 | 15.852 |
| 92 | 140 | 232 | 5.40 | 980.9926 | 618.870 | 16.391 |
| 92 | 142 | 234 | 5.80 | 978.1897 | 620.498 | 16.340 |
| 92 | 144 | 236 | 5.75 | 975.4186 | 618.472 | 16.195 |
| 92 | 146 | 238 | 5.90 | 972.6787 | 618.055 | 16.093 |
| 92 | 148 | 240 | 5.80 | 969.9693 | 615.672 | 15.942 |
| 94 | 144 | 238 | 5.30 | 1015.6662 | 638.268 | 16.620 |
| 94 | 146 | 240 | 5.50 | 1012.8370 | 638.326 | 16.529 |
| 94 | 148 | 242 | 5.50 | 1010.0391 | 636.691 | 16.395 |
| 94 | 150 | 244 | 5.30 | 1007.2718 | 633.371 | 16.221 |
| 94 | 152 | 246 | 5.30 | 1004.5347 | 631.773 | 16.092 |
| 96 | 146 | 242 | 5.00 | 1053.7127 | 657.730 | 16.937 |
| 96 | 148 | 244 | 5.00 | 1050.8258 | 656.054 | 16.801 |
| 96 | 150 | 246 | 4.70 | 1047.9703 | 651.594 | 16.597 |
| 96 | 152 | 248 | 5.00 | 1045.1455 | 652.755 | 16.537 |
| 96 | 154 | 250 | 4.40 | 1042.3510 | 645.431 | 16.264 |
| 98 | 154 | 252 | 4.80 | 1083.5862 | 673.167 | 16.873 |
| | | | | | **Average value of $a_s$** | **16.481** |

Corresponding author e-mail: serkan.akkoyun@gmail.com

We have tested semi-empirical mass formula (Eq. 2.1) with Krane coefficient ($a_v$=15.5, $a_s$=16.8, $a_C$=0.72, $a_a$=23 and $a_p$=34). The mean square error (MSE) value between theoretical masses of the nuclei and the experimental masses has been obtained 100.9 for 3245 isotopes from A=20 to 295. If we use the new surface coefficient as $a_s$=16.481 in same formula, the MSE value has been achieved as 29.6 which gives 3.4 factor better result than Krane surface coefficient gives. In Fig. 1.a, the differences between experimental binding energies ($BE_{exp}$) and theoretical binding energies ($BE_{theo}$) calculated by Krane coefficients have been shown. The deviations from experimental values are lied between about -10 to 40 MeV. In Fig. 1.b, we have also shown these differences by redefined surface coefficient. The deviations are lied between about -10 to 20 MeV.

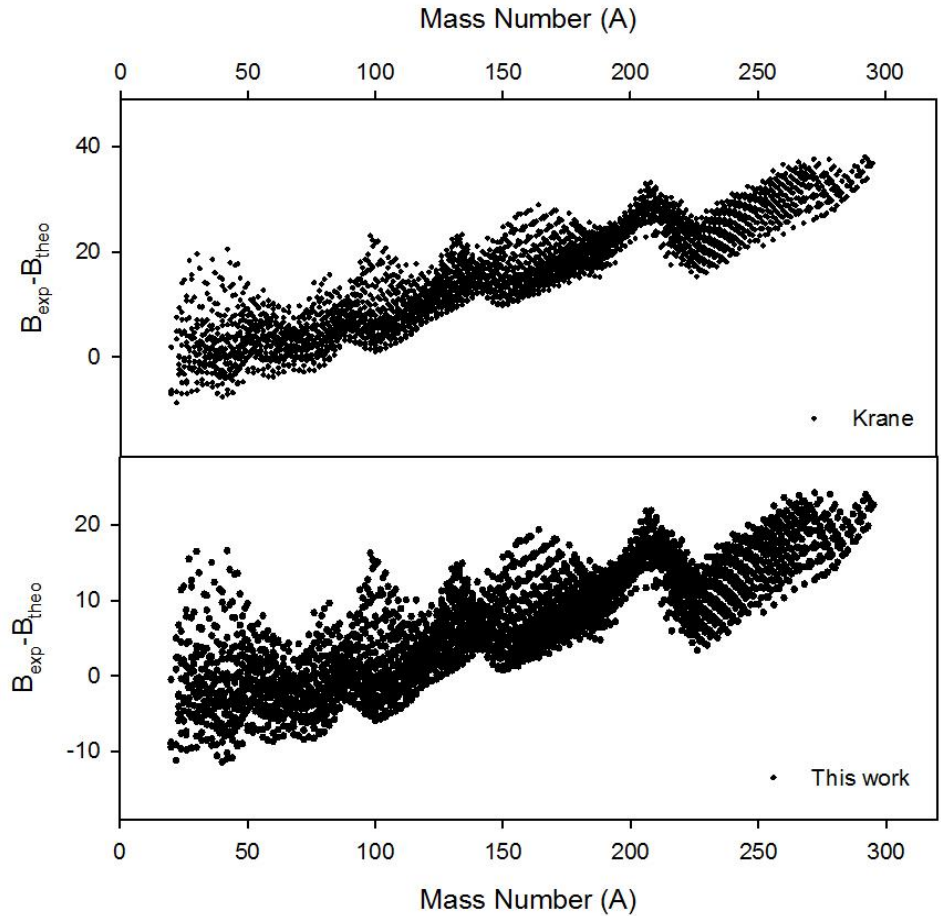

**Fig.1** The difference between experimental and theoretical binding energies with Krane coefficient (a) and the coefficient obtained in this work (b)

Corresponding author e-mail: serkan.akkoyun@gmail.com

## 4. Conclusion

In this work, the experimental fission barrier energy values exist for 36 nuclei have been used for determination of the surface energy coefficient in semi-empirical mass formula of liquid drop model of the nucleus. We have considered conventional mass formula with Krane coefficient. We have borrowed Coulomb energy coefficient as existing value and calculated surface energy terms and then their coefficients for each 36 nuclei. After obtaining the coefficients, we have calculated the average value. The redefined value of the surface coefficient is $a_s$=16.481. Other coefficients remain same, when we used this coefficient in semi-empirical formula, the result is 3.4 factor better the result of Krane surface coefficient.

**Acknowledge**

This work was supported by Cumhuriyet University Scientific Research Center with project number SHMYO-008.

Corresponding author e-mail: serkan.akkoyun@gmail.com

Corresponding author e-mail: serkan.akkoyun@gmail.com